\documentclass{llncs}
\usepackage{amsmath}
\usepackage{amssymb}
\usepackage{enumerate}
\usepackage{amsfonts}
\usepackage{graphicx}
\usepackage{xspace}
\usepackage{mathrsfs}

\newcounter{instr}

\newcommand{\tb}[1]{\textbf{#1}}
\newcommand{\bigO}{\mathcal{O}}

\def\pp{\mathinner{\ldotp\ldotp}}

\renewcommand{\t}{t\xspace}
\newcommand{\p}{p\xspace}
\renewcommand{\tb}{T\xspace}
\newcommand{\pb}{P\xspace}
\newcommand{\epsma}{EPSM$a$\xspace}
\newcommand{\epsmb}{EPSM$b$\xspace}
\newcommand{\epsmc}{EPSM$c$\xspace}

\title{Fast Packed String Matching for Short Patterns}

\author{Simone Faro$^\dag$ \and M. O\u{g}uzhan K\"{u}lekci$^\ddag$}
\institute{
$^\dag$Dipartimento di Matematica e Informatica, Universit\`a di Catania, Italy\\
$^\ddag$T\"UB\.ITAK National Research Institute of Electronics and Cryptology, Turkey\\
\email{faro@dmi.unict.it, oguzhan.kulekci@tubitak.gov.tr
}
}

\begin{document}

\maketitle

\begin{abstract}
Searching for all occurrences of a pattern in a text is a fundamental problem in computer science with applications in many other fields, like natural language processing, information retrieval and computational biology.  
In the last two decades a general trend has appeared trying to exploit the power of the word RAM model to speed-up the performances of classical string matching algorithms. In this model an algorithm operates on words of length $w$, grouping blocks of characters, and arithmetic and logic operations on the words take one unit of time.

In this paper we use specialized word-size packed string matching instructions, based on the Intel streaming SIMD
extensions (SSE) technology, to design very fast string matching algorithms in the case of short patterns. From our
experimental results it turns out that, despite their quadratic worst case time complexity, the new presented algorithms
become the clear winners on the average for short patterns, when compared against the most effective algorithms known in
literature. 
\end{abstract}

\section{Introduction}
Given a text $\t$ of length $n$ and a pattern $\p$ of length $m$ over some alphabet $\Sigma$ of size $\sigma$, the
\emph{exact string matching problem} consists in finding \emph{all} occurrences of
the pattern $\p$ in $\t$. This problem has been extensively studied in computer science 
because
of its direct application to many areas. Moreover string matching algorithms are basic components in many software
applications and play an important role in theoretical computer science by providing challenging problems.

In a computational model where the matching algorithm is restricted to read all the characters of the text one by one the
optimal complexity is $\bigO(n)$, and was achieved the first time by the well known Knuth-Morris-Pratt algorithm
\cite{knuth77} (KMP).
However in many practical cases it is possible to avoid reading all the characters of the text achieving sub-linear
performances on average. The optimal average $\bigO(\frac{n\log_{\sigma}m}{m})$ time complexity \cite{yao79} was reached for
the first time by the Backward-DAWG-Matching algorithm \cite{crochemore94} (BDM).  
However, most of the algorithms with a sub-linear average behavior may have to read all the text characters in the worst
case. 
It is interesting to note that many of those algorithms have an even worse $\bigO(nm)$-time complexity in the
worst-case \cite{charras04}.

In the last two decades a lot of work has been made in order to exploit the power of the word RAM model of computation to speed-up classical string matching algorithms.
In this model, the computer operates on words of length $w$, thus blocks of characters are read and processed at once. This means that usual arithmetic and logic operations on the words all take one unit of time.

Most of the solutions which exploit the word RAM model are based on the \emph{bit-parallelism} technique or on the \emph{packed string matching} technique.  

The \emph{bit-parallelism} technique \cite{baezayates92} takes advantage of the intrinsic parallelism of the bit operations
inside a computer word, allowing to cut down the number of operations that an algorithm performs by a factor up to $w$.
Bit-parallelism is particularly suitable for the efficient simulation of nondeterministic
automaton. The first algorithm based on it, named Shift-Or \cite{baezayates92} (SO),  simulates efficiently the
nondeterministic version of the {KMP} automaton and runs in $\bigO(n\lceil \frac{m}{w}\rceil)$, which is still
considered among the
best practical algorithms in the case of very short patterns and small alphabets \cite{faro13,faro10}.
Later a very fast BDM-like algorithm (BNDM), based on the bit-parallel simulation of the nondeterministic suffix
automaton, was presented in \cite{navarro98}. Some variants of the BNDM algorithm \cite{faro08,faro09,durian09,tarhio11} are among the most practical efficient
solutions in literature (see \cite{faro13,faro10}).
However, the bit-parallel encoding requires one bit per pattern symbol, for a total of $\lceil \frac{m}{w}\rceil$ computer
words. Thus, as long as a pattern fits in a computer word, bit-parallel algorithms are extremely fast,
otherwise their performances degrades considerably as $\lceil \frac{m}{w}\rceil$ grows. 
Though there are a few techniques to maintain good performance in the case of long patterns  \cite{kulekci10,DPST10,CFG10}, such limitation is intrinsic.

In the \emph{packed string matching}  technique multiple characters are packed into one larger word, so that the
characters can be compared in bulk rather than individually. In this context, if the characters of a string are drawn
from an alphabet of size $\sigma$, then $\lfloor \frac{w}{\log \sigma} \rfloor$ different characters fit in a single word, using
$\lfloor \log \sigma \rfloor$ bits per characters. The packing factor is $\alpha= \frac{w}{\log \sigma}$.

A first theoretical result in packed string matching was proposed by Fredriksson \cite{fredriksson02}. 
He presented a general scheme that can be applied to speed-up many pattern matching algorithms. 
His approach relies on the use of the \emph{four russian} technique (i.e. tabulation), achieving in favorable cases a
$\bigO(n^{\varepsilon}m)$-space and $\bigO( \frac{n} {m \log \sigma} + n^{\varepsilon}m + occ)$-time complexity, where  $\varepsilon > 0$ denotes an arbitrary small constant, and $occ$ denotes the number of occurrences of $\p$ in $\t$. 
Bille \cite{bille11} presented an alternative solution with  $\bigO(\frac{n}{\log_{\sigma}n}+m +occ)$-time and
$\bigO(n^{\varepsilon}+m)$-space complexities by an efficient segmentation and coding of the KMP automaton.
Recently Belazzougui \cite{belazzougui10} proposed a packed string matching algorithm which works in $\bigO(\frac{n}{m} + \frac{n}{\alpha}+m+occ)$ time and $\bigO(m)$ space, reaching the optimal $\bigO(\frac{n}{\alpha}+occ)$-time bound for  $\alpha \leq m \leq \frac{n}{\alpha}$. 
However, none of these results is of any practical interest.

The first algorithm that achieves good practical and theoretical results was very recently proposed by Ben-Kiki
\emph{et al.} \cite{benkiki11}. The algorithm is based on two specialized packed string instructions, 
the \textsf{pcmpestrm} and the 
\textsf{pcmpestri} instructions \cite{Intell11},
and reaches the optimal $\bigO(\frac{n}{\alpha}+occ)$-time complexity requiring only $\bigO(1)$ extra space. Moreover the authors showed  
that their algorithm turns out to be among the fastest string matching solutions in the case
of very short patterns.
However, it has to be noticed that on current generation Intel Sandy Bridge processors, \textsf{pcmpestrm} and
\textsf{pcmpestri} have 2-cycle throughput and 7- and 8-cycle latency, respectively \cite{Intell11}. 

When the length of the searched pattern increases, another algorithm named Streaming {SIMD} Extensions Filter (SSEF),
presented by K\"{u}lekci in \cite{kulekci09} (and extended to multiple pattern matching in \cite{FK12}), exploits the advantages of the word-RAM model. 
Specifically  it uses a filter method that inspects blocks of characters instead of reading them one by one. Despite 
its $\bigO(nm)$ worst case time complexity, the SSEF algorithm turns out to be among the fastest solutions when
searching for long patterns \cite{faro13,faro10}.
Efficient solutions have been also designed for searching on packed DNA sequences~\cite{RTT02,FL09}. However in this paper we do not take into account this type of solutions since they require a different type of data representation.


Streaming SIMD technology offers single-instructions to perform a variety of tests on packed strings.
Unfortunately those instructions are heavier than other instructions provided in the same family as a consequence of
their relatively high latencies. Hence, in this paper we focus on design of algorithms using instructions with low latency and
throughput, when compared with those used in \cite{benkiki11}.
Specifically  we present a new practical and efficient algorithm for the exact packed string matching problem that
turns out to be faster than the best algorithms known in literature in the case of short patterns. 
The algorithm, named Exact Packed String Matching (EPSM), is based on three different search procedures used for,
respectively, very short patterns ($0<m<\frac{\alpha}{2}$), short patterns ($\frac{\alpha}{2} \leq m < \alpha$) and
medium length patterns ($m\geq \alpha$). 
They use  specialized packed string instructions with a low latency and
throughput, if compared with those used in \cite{benkiki11}.
All search procedures have an $\bigO(nm)$ worst case time complexity. However, they have very good performances on
average. 
In the case of very short patterns, i.e. when $m\leq \frac{\alpha}{2}$, the first two search procedures achieve, respectively, a $\bigO(n+occ)$ and an optimal $\bigO(\frac{n}{\alpha} + occ)$-time complexity.


The paper is organized as follows.
In Section~\ref{sec:notions}, we introduce some notions and terminologies.
We then present a new algorithm for the packed string matching problem in Section~\ref{sec:new} and report experimental results on short patterns in Section~\ref{sec:results}. 
Conclusions  are given in Section~\ref{sec:conclusions}.

\section{Notions and Terminology} \label{sec:notions}
Throughout the paper we will make use of the following notations and
terminology.  A string $\p$ of length $m> 0$ is represented as a finite array $\p[0\pp m-1]$ of characters from a finite alphabet $\Sigma$ of size $\sigma$. 
Thus $\p[i]$ will denote the $(i+1)$-st character of $\p$, for $0\leq i < m$, and $\p[i\pp j]$ will denote the \emph{factor} (or \emph{substring}) of $\p$ contained between the $(i+1)$-st and the $(j+1)$-st characters of
$\p$, for $0\leq i \leq j < m$. In some cases we will denote by $\p_i$ the $(i+1)$-st character of $\p$, so that $\p_i = \p[i]$ and $\p = \p_0\p_1\ldots\p_{m-1}$.

We indicate with symbol $w$ the number of bits in a computer word and with symbol $\gamma = \lceil \log\sigma \rceil$
the number of bits used for encoding a single character of the alphabet $\Sigma$.  The number of characters of the
alphabet that fit in a single word is shown by $\alpha = \lfloor w/\gamma \rfloor$. Without lose in generality
we will assume along the paper that $\gamma$ divides $w$ and that $\alpha$ is an even value.


In chunks of $\alpha$ characters, the string $\p$ is represented by an array $P[0\pp k-1]$ of length
$k=(m-1)/\alpha+1$. In particular we denote $\pb = \pb_0\pb_1\pb_2\ldots \pb_{k-1}$, where $\pb_i =
\p_{i\alpha}\p_{i\alpha+1}\p_{i\alpha+2} \ldots \p_{i\alpha+\alpha-1}$, for $0 \leq i < k$. The last block $\pb_{k-1}$
is not complete if $m \bmod \alpha \neq 0$. In that case, the rightmost remaining characters of the block are set to
zero.

Although different values of $\alpha$ and $\gamma$ are possible, in most cases we assume that $\alpha=16$ and
$\gamma=8$, which is the most common case when working with characters in ASCII code and in a word RAM model with
128-bit registers, which are almost all available in recent commodity processors supporting single
instruction multiple data (SIMD) operations.

Finally, we recall the notation of some bitwise infix operators on computer words, namely the bitwise \texttt{and} ``$\&$'', the bitwise \texttt{or} ``$|$'' and the \texttt{left shift} ``$\ll$'' operator (which shifts to the left its first argument by a number of bits equal to its second argument).

\section{A New Packed String Matching Algorithm} \label{sec:new}

In this section we present a new packed string matching algorithm, named Exact Packed String Matching algorithm (EPSM),
which turns out to be efficient in the case of short patterns.  
{EPSM} is based on three different auxiliary algorithms, which we name \epsma, \epsmb and \epsmc, respectively.

The first two auxiliary algorithms are designed to search for patterns of length, at most, $\alpha/2$. 
When the length of the pattern is longer than $\alpha/2$ the algorithms adopt a filter mechanism:  they first search for a substring of the pattern of length $\alpha/2$ and, when a candidate occurrence has been found, a naive check follows.
The third algorithm adopts a filtering based solution.
 
All three algorithms run in $\bigO(nm)$ worst case time complexity and use, respectively,  $\bigO(\min\{m,\alpha\})$, $\bigO(1)$ and $\bigO(2^k)$ additional space, where $k$ is a constant parameter. 
However, when $m\leq \alpha/2$ the \epsma and \epsmb algorithms reach, respectively, an $\bigO(m \alpha + \frac{mn}{\alpha} + occ)$ and $\bigO(\frac{n}{\alpha}+occ)$ time complexity.
The first search procedure is designed to be extremely fast in the case of very short patterns, i.e.
when $m\leq \frac{\alpha}{2}$, the second algorithm turns out to be a good choice when $\frac{\alpha}{2} \leq m <
\alpha$, while the third algorithm turns out to be effective when $m \geq \alpha$. In practical cases we tuned the EPSM
algorithm in order to run \epsma when $0< m<4$, \epsmb when $4\leq m < 16$, and to run \epsmc in all other cases.

In what follows, we first describe in Section \ref{sec:model} the computational model we use for the description of our
solutions. Then  we independently  present the three auxiliary algorithms \epsma, in Section \ref{sec:new1}, \epsmb, in
Section \ref{sec:new2}, and \epsmc in Section \ref{sec:new3}.

\subsection{The Model}\label{sec:model}
In the design of our algorithms we use specialized word-size packed string matching instructions, based on the Intel
streaming SIMD extensions (SSE) technology.
SIMD instructions exist in many recent microprocessors supporting parallel execution of some operations on multiple data simultaneously via a set of special instructions working on limited number of special registers. Although the usage of SIMD is explored deeply in multimedia processing, implementation of encryption/decryption algorithms, and on some scientific calculations, it has not been much addressed in pattern matching.

In our model of computation we suppose that $w$ is the number of bits in a word and $\sigma$ is the size of the alphabet. We indicate with the symbol $\alpha= \frac{w}{\log \sigma}$ the number of characters which fit in a single computer word.

In most practical applications we have $\sigma = 256$ (ASCII code). Moreover SSE specialized instructions allow to work on 128-bit registers, thus reading and processing blocks of sixteen 8-bit characters in a single time unit (thus $\alpha=16$). 

In the design of our algorithms we make use of the following specialized word-size packed instructions. For each
instruction we describe how it could be emulated by using SSE specialized intrinsics.

\subsubsection{\textsf{wscmp}$(a,b)$} \textbf{(\emph{word-size compare instruction})}\\
Compares two $w$-bit words, handled as a block of $\alpha$  characters.
In particular if $a=a_0a_1\ldots a_{\alpha-1}$ and $b=b_0b_1\ldots b_{\alpha-1}$ are the two $w$-bit integer parameters, \textsf{wscmp}$(a,b)$ returns an $\alpha$-bit value $r=r_0r_1\ldots r_{\alpha-1}$, where $r_i = 1$ if and only if $a_i=b_i$, and $r_i = 0$ otherwise.
Below we give an example of the application of \textsf{wscmp}$(a,b)$, assuming $w=48$, $\gamma=4$ and $\alpha=12$.
\begin{center}
\setlength{\unitlength}{0.01\textwidth}
\setlength{\fboxrule}{1mm}
\begin{picture}(100,26)
 \put(4,16){ \makebox(4,4)[c]{$a$:}}
 \put(8,16){ \framebox(84,4)[c]{}}
 \put(8,16){  \makebox(7,4)[c]{0110.} }
 \put(15,16){  \makebox(7,4)[c]{0010.} }
 \put(22,16){  \makebox(7,4)[c]{0111.} }
 \put(29,16){  \makebox(7,4)[c]{1010.} }
 \put(36,16){  \makebox(7,4)[c]{0010.} }
 \put(43,16){  \makebox(7,4)[c]{1110.} }
 \put(50,16){  \makebox(7,4)[c]{0010.} }
 \put(57,16){  \makebox(7,4)[c]{0100.} }
 \put(64,16){  \makebox(7,4)[c]{0110.} }
 \put(71,16){  \makebox(7,4)[c]{0111.} }
 \put(78,16){  \makebox(7,4)[c]{0100.} }
 \put(85,16){  \makebox(7,4)[c]{0010} }
 \put(8,20){  \makebox(7,4)[c]{$_0$} }
 \put(15,20){  \makebox(7,4)[c]{$_1$} }
 \put(22,20){  \makebox(7,4)[c]{$_2$} }
 \put(29,20){  \makebox(7,4)[c]{$_3$} }
 \put(36,20){  \makebox(7,4)[c]{$_4$} }
 \put(43,20){  \makebox(7,4)[c]{$_5$} }
 \put(50,20){  \makebox(7,4)[c]{$_6$} }
 \put(57,20){  \makebox(7,4)[c]{$_7$} }
 \put(64,20){  \makebox(7,4)[c]{$_8$} }
 \put(71,20){  \makebox(7,4)[c]{$_9$} }
 \put(78,20){  \makebox(7,4)[c]{$_{10}$} }
 \put(85,20){  \makebox(7,4)[c]{$_{11}$} }

 \put(4,10){ \makebox(4,4)[c]{$b$:}}
 \put(8,10){ \framebox(84,4)[c]{}}
 \put(8,10){  \makebox(7,4)[c]{0100.} }
 \put(15,10){  \makebox(7,4)[c]{0010.} }
 \put(22,10){  \makebox(7,4)[c]{0000.} }
 \put(29,10){  \makebox(7,4)[c]{0111.} }
 \put(36,10){  \makebox(7,4)[c]{1111.} }
 \put(43,10){  \makebox(7,4)[c]{0010.} }
 \put(50,10){  \makebox(7,4)[c]{0010.} }
 \put(57,10){  \makebox(7,4)[c]{1100.} }
 \put(64,10){  \makebox(7,4)[c]{0110.} }
 \put(71,10){  \makebox(7,4)[c]{0100.} }
 \put(78,10){  \makebox(7,4)[c]{1110.} }
 \put(85,10){  \makebox(7,4)[c]{0010} }

 \put(4,4){ \makebox(4,4)[c]{$r$:}}
 \put(8,4){ \framebox(84,4)[c]{}}
 \put(8,4){  \makebox(7,4)[c]{0} }
 \put(15,4){  \makebox(7,4)[c]{1} }
 \put(22,4){  \makebox(7,4)[c]{0} }
 \put(29,4){  \makebox(7,4)[c]{0} }
 \put(36,4){  \makebox(7,4)[c]{0} }
 \put(43,4){  \makebox(7,4)[c]{0} }
 \put(50,4){  \makebox(7,4)[c]{1} }
 \put(57,4){  \makebox(7,4)[c]{0} }
 \put(64,4){  \makebox(7,4)[c]{1} }
 \put(71,4){  \makebox(7,4)[c]{0} }
 \put(78,4){  \makebox(7,4)[c]{0} }
 \put(85,4){  \makebox(7,4)[c]{1} }

 \put(15,3){ \framebox(7,18)[c]{}}
 \put(50,3){ \framebox(7,18)[c]{}}
 \put(64,3){ \framebox(7,18)[c]{}}
 \put(85,3){ \framebox(7,18)[c]{}}
\end{picture}
\end{center}

The \textsf{wscmp} specialized instruction can be emulated in constant time by using the following sequence of specialized SIMD instructions

\begin{quote}
	\textsf{$h \leftarrow$ \_mm\_cmpeq\_epi8$(a,b)$} \\
        	\textsf{$r \leftarrow$  \_mm\_movemask\_epi8$(h)$}
\end{quote}

Specifically the \textsf{\_mm\_cmpeq\_epi8} instruction compares two 128-bit words, handled as a block of sixteen 8-bit
values,  and returns a 128-bit value $h=h_0h_1\ldots h_{15}$, where $h_i = 1^8$ if and only if $a_i=b_i$, and $h_i = 0^8$
otherwise. It has a 0.5-cycle throughput and a 1-cycle latency\\
The \textsf{\_mm\_movemask\_epi8} instruction gets a 128 bit parameter $h$, handled as sixteen 8-bit integers, and creates a 16-bit mask from the most significant bits of the 16 integers in $h$, and zero extends the upper bits.

\subsubsection{\textsf{wsmatch}$(a,b)$} \textbf{(\emph{word-size pattern matching instruction})}\\
reports all occurrences of a short string $b$ in a $w$-bit parameter $a$, handled as a string of $\alpha$ characters.
The parameter $b$ is a string of length $k\leq \alpha$.\\
Specifically, if $a=a_0a_1\ldots a_{\alpha-1}$, and $b=b_0b_1\ldots b_{k-1}$,   then the \textsf{wsmatch}$(a,b)$ instruction returns an $\alpha$-bit integer value, $r=r_0r_1\ldots r_{\alpha-1}$, where $r_i=1$ if and only if $a_{i+j} = b_j$ for $j=0\ldots k-1$, i.e. an occurrence of $b$ in $a$ begins at position $i$. Notice that $r_i=0$ for $\alpha-k<i<\alpha$, since no occurrence of $b$ in $a$ could begin at a position greater than $\alpha-k$.
Below we give an example of the application of \textsf{wsmatch}$(a,b)$, assuming $w=48$, $\gamma=4$, $\alpha=12$ and $k=3$.
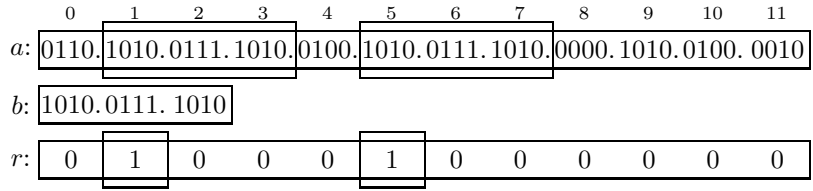
\begin{center}
\setlength{\unitlength}{0.01\textwidth}
\setlength{\fboxrule}{1mm}
\begin{picture}(100,26)
 \put(4,16){ \makebox(4,4)[c]{$a$:}}
 \put(8,16){ \framebox(84,4)[c]{}}
 \put(8,16){  \makebox(7,4)[c]{0110.} }
 \put(15,16){  \makebox(7,4)[c]{1010.} }
 \put(22,16){  \makebox(7,4)[c]{0111.} }
 \put(29,16){  \makebox(7,4)[c]{1010.} }
 \put(36,16){  \makebox(7,4)[c]{0100.} }
 \put(43,16){  \makebox(7,4)[c]{1010.} }
 \put(50,16){  \makebox(7,4)[c]{0111.} }
 \put(57,16){  \makebox(7,4)[c]{1010.} }
 \put(64,16){  \makebox(7,4)[c]{0000.} }
 \put(71,16){  \makebox(7,4)[c]{1010.} }
 \put(78,16){  \makebox(7,4)[c]{0100.} }
 \put(85,16){  \makebox(7,4)[c]{0010} }
 \put(8,20){  \makebox(7,4)[c]{$_0$} }
 \put(15,20){  \makebox(7,4)[c]{$_1$} }
 \put(22,20){  \makebox(7,4)[c]{$_2$} }
 \put(29,20){  \makebox(7,4)[c]{$_3$} }
 \put(36,20){  \makebox(7,4)[c]{$_4$} }
 \put(43,20){  \makebox(7,4)[c]{$_5$} }
 \put(50,20){  \makebox(7,4)[c]{$_6$} }
 \put(57,20){  \makebox(7,4)[c]{$_7$} }
 \put(64,20){  \makebox(7,4)[c]{$_8$} }
 \put(71,20){  \makebox(7,4)[c]{$_9$} }
 \put(78,20){  \makebox(7,4)[c]{$_{10}$} }
 \put(85,20){  \makebox(7,4)[c]{$_{11}$} }

 \put(4,10){ \makebox(4,4)[c]{$b$:}}
 \put(8,10){ \framebox(21,4)[c]{}}
 \put(8,10){  \makebox(7,4)[c]{1010.} }
 \put(15,10){  \makebox(7,4)[c]{0111.} }
 \put(22,10){  \makebox(7,4)[c]{1010} }
 \put(15,15){ \framebox(21,6)[c]{}}
 \put(43,15){ \framebox(21,6)[c]{}}

 \put(4,4){ \makebox(4,4)[c]{$r$:}}
 \put(8,4){ \framebox(84,4)[c]{}}
 \put(8,4){  \makebox(7,4)[c]{0} }
 \put(15,4){  \makebox(7,4)[c]{1} }
 \put(22,4){  \makebox(7,4)[c]{0} }
 \put(29,4){  \makebox(7,4)[c]{0} }
 \put(36,4){  \makebox(7,4)[c]{0} }
 \put(43,4){  \makebox(7,4)[c]{1} }
 \put(50,4){  \makebox(7,4)[c]{0} }
 \put(57,4){  \makebox(7,4)[c]{0} }
 \put(64,4){  \makebox(7,4)[c]{0} }
 \put(71,4){  \makebox(7,4)[c]{0} }
 \put(78,4){  \makebox(7,4)[c]{0} }
 \put(85,4){  \makebox(7,4)[c]{0} }
 \put(15,3){ \framebox(7,6)[c]{}}Ä
 \put(43,3){ \framebox(7,6)[c]{}}
\end{picture}
\end{center}

The \textsf{wsmatch}$(a,b)$ instruction  can be emulated in constant time by using the following sequence of SIMD specialized instructions
\begin{quote}
	\textsf{$h  \leftarrow$ \_mm\_mpsadbw\_epu8$(a,b)$}\\
        	\textsf{$\ell \leftarrow$ \_mm\_cmpeq\_epi8$(h,z)$}\\
        	\textsf{$r \leftarrow$  \_mm\_movemask\_epi8$(\ell)$}
\end{quote}
where $z$ is a $128$-bit register with all bits set to $0$, i.e. $z=0^{128}$. 

Specifically the \textsf{\_mm\_mpsadbw\_epu8$(a,b)$} instruction gets  two $128$-bit words, handled as a block of
sixteen 8-bit values,  and returns a $128$-bit value $r=r_0r_1\ldots r_{7}$, where $r_i$ is computed as
$
	r_i = \sum_{j=0}^{4} | a_{i+j} - b_j |
$
for $i=0\ldots 7$. Thus we have that $r_i=0^{16}$ if and only if $a_{i+j} = b_j$ for $j=0\ldots 4$, i.e. an occurrence
of the prefix of $b$ with length $4$   begins in $a$ at position $i$. The \textsf{\_mm\_mpsadbw\_epu8} instruction has
1-cycle throughput and a 4-cycle latency.
The  \textsf{\_mm\_cmpeq\_epi8}  and \textsf{\_mm\_movemask\_epi8} instructions have been described above.

\subsubsection{\textsf{wsblend}$(a,b)$} \textbf{(\emph{word-size blend instruction})}\\
blends two $w$-bit parameters, handled as two blocks of $\alpha$ characters.
Specifically if $a=a_0a_1\ldots a_{\alpha-1}$ and $b=b_0b_1\ldots b_{\alpha-1}$, the instruction  returns a $w$-bit integer $r=r_0r_1\ldots r_{\alpha-1}$, where $r_i = a_{i+\alpha/2}$, if $0\leq i <\alpha/2$, and $r_i=b_{i-\alpha/2}$ if $\alpha/2\leq i<\alpha$, i.e.  $r=a_{\frac{\alpha}{2}}a_{\frac{\alpha}{2}+1}\ldots a_{\alpha-1}b_0b_1\ldots b_{\frac{\alpha}{2}-1}$.
Below we give an example of the application of \textsf{wsmatch}$(a,b)$, assuming $w=48$, $\gamma=4$ and $\alpha=12$.
\begin{center}
\setlength{\unitlength}{0.01\textwidth}
\setlength{\fboxrule}{1mm}
\begin{picture}(100,26)
 \put(4,16){ \makebox(4,4)[c]{$a$:}}
 \put(8,16){ \framebox(84,4)[c]{}}
 \put(8,16){  \makebox(7,4)[c]{0110.} }
 \put(15,16){  \makebox(7,4)[c]{0010.} }
 \put(22,16){  \makebox(7,4)[c]{0111.} }
 \put(29,16){  \makebox(7,4)[c]{1010.} }
 \put(36,16){  \makebox(7,4)[c]{0010.} }
 \put(43,16){  \makebox(7,4)[c]{1110.} }
 \put(50,16){  \makebox(7,4)[c]{0010.} }
 \put(57,16){  \makebox(7,4)[c]{0100.} }
 \put(64,16){  \makebox(7,4)[c]{0110.} }
 \put(71,16){  \makebox(7,4)[c]{0111.} }
 \put(78,16){  \makebox(7,4)[c]{0100.} }
 \put(85,16){  \makebox(7,4)[c]{0010} }
 \put(8,20){  \makebox(7,4)[c]{$_0$} }
 \put(15,20){  \makebox(7,4)[c]{$_1$} }
 \put(22,20){  \makebox(7,4)[c]{$_2$} }
 \put(29,20){  \makebox(7,4)[c]{$_3$} }
 \put(36,20){  \makebox(7,4)[c]{$_4$} }
 \put(43,20){  \makebox(7,4)[c]{$_5$} }
 \put(50,20){  \makebox(7,4)[c]{$_6$} }
 \put(57,20){  \makebox(7,4)[c]{$_7$} }
 \put(64,20){  \makebox(7,4)[c]{$_8$} }
 \put(71,20){  \makebox(7,4)[c]{$_9$} }
 \put(78,20){  \makebox(7,4)[c]{$_{10}$} }
 \put(85,20){  \makebox(7,4)[c]{$_{11}$} }
 \put(50,15){ \framebox(42,6)[c]{}}
 \put(4,10){ \makebox(4,4)[c]{$b$:}}
 \put(8,10){ \framebox(84,4)[c]{}}
 \put(8,10){  \makebox(7,4)[c]{0100.} }
 \put(15,10){  \makebox(7,4)[c]{0010.} }
 \put(22,10){  \makebox(7,4)[c]{0000.} }
 \put(29,10){  \makebox(7,4)[c]{0111.} }
 \put(36,10){  \makebox(7,4)[c]{1111.} }
 \put(43,10){  \makebox(7,4)[c]{0010.} }
 \put(50,10){  \makebox(7,4)[c]{0010.} }
 \put(57,10){  \makebox(7,4)[c]{1100.} }
 \put(64,10){  \makebox(7,4)[c]{0110.} }
 \put(71,10){  \makebox(7,4)[c]{0100.} }
 \put(78,10){  \makebox(7,4)[c]{1110.} }
 \put(85,10){  \makebox(7,4)[c]{0010} }
 \put(8,9){ \framebox(42,6)[c]{}}

 \put(4,4){ \makebox(4,4)[c]{$r$:}}
 \put(8,4){ \framebox(84,4)[c]{}}
 \put(8,4){  \makebox(7,4)[c]{0010.} }
 \put(15,4){  \makebox(7,4)[c]{0100.} }
 \put(22,4){  \makebox(7,4)[c]{0110.} }
 \put(29,4){  \makebox(7,4)[c]{0111.} }
 \put(36,4){  \makebox(7,4)[c]{0100.} }
 \put(43,4){  \makebox(7,4)[c]{0010.} }

 \put(50,4){  \makebox(7,4)[c]{0100.} }
 \put(57,4){  \makebox(7,4)[c]{0010.} }
 \put(64,4){  \makebox(7,4)[c]{0000.} }
 \put(71,4){  \makebox(7,4)[c]{0111.} }
 \put(78,4){  \makebox(7,4)[c]{1111.} }
 \put(85,4){  \makebox(7,4)[c]{0010} }
\end{picture}
\end{center}

The \textsf{wsblend}$(a,b)$ instruction  can be emulated in constant time by using the following sequence of SIMD specialized instructions
\begin{quote}
	\textsf{$h  \leftarrow$ \_mm\_blend\_epi16$(a, b, c)$}\\
        	\textsf{$r \leftarrow$ \_mm\_shuffle\_epi32$(h$, \_MM\_SHUFFLE$(1,0,3,2))$}
\end{quote}
Such  instruction blends two 128-bit integers, $a=a_0a_1\ldots a_{7}$ and $b=b_0b_1\ldots b_{7}$, handled as packed 16-bit integers, according to a third parameter $c$. In particular it returns a 128-bit integer $r=r_0r_1\ldots r_7$ where $r_i=a_i$ if $c_i = 0$, and $r_i=b_i$ otherwise. If we set $c=0^{64}1^{64}$ we get $r = a_0a_1a_2a_3b_4b_5b_6b_7$. The \textsf{\_mm\_blend\_epi16} instruction has 0.5-cycle throughput and a 1-cycle latency.\\
The \textsf{\_mm\_shuffle\_epi32}  instruction shuffles a $w$-bit parameter, $a=a_0a_1a_2a_3$, handled as four 32-bit values, according to the order of the \textsf{\_MM\_SHUFFLE} macro. In this case we get $r=a_2a_3a_0a_1$. The \textsf{\_mm\_shuffle\_epi32} instruction has 1-cycle throughput and a 1-cycle latency.

\subsubsection{\textsf{wscrc}$(a)$} \textbf{(\emph{word-size cyclic redundancy check instruction})}\\
computes the $32$-bit cyclic redundancy checksum (CRC) signature for a $w$-bit parameter.
It is an error-detecting code commonly used in digital networks and storage devices to detect accidental changes to raw data and  can also be used as a hash function. 

The \textsf{wscrc}$(a)$ instruction  can be emulated in constant time by using the \textsf{\_mm\_crc32\_u64}$(a)$ SIMD specialized instructions,
 which computes the $32$ bit cyclic redundancy check of a $64$-bit block according to a polynomial.
Such instruction has a 1-cycle throughput and a 3-cycle latency, thus provides a robust and fast way of computing hash values.

\subsubsection{Additional specialized instructions}~~\\
In addition to the above listed instructions, given an $\alpha$-bit register $r$, in our description we make use of the symbol $\{r\}$ to indicate the set of bits in $r$ whose value is set.
More formally, given an $\alpha$-bit register $r=r_0r_1r_2 \ldots r_{\alpha-1}$, we have
$
	\{r\} = \{ i\ |\ 0\leq i< \alpha \textrm{ and } r_i=1\}
$.
Moreover, given a value $s\in \mathbb{N}$, we use for simplicity the expression $s+\{r\}$ to indicate the set of values $\{s+i\ |\ i\in\{r\}\}$.

The cardinality of the set $\{r\}$ can be computed in constant time by using the SIMD specialized instructions \textsf{\_mm\_popcnt\_u32}$(r)$ which calculates the number of bits of the parameter $r$ that are set to $1$. Such instruction has 1-cycle throughput and a 3-cycle latency.

Differently the list of values in $\{r\}$ can be efficiently listed in $\bigO(\alpha)$-time and $\bigO(1)$-space, or using a tabulation approach, in $\bigO(|\{r\}|)$-time and $\bigO(2^{\alpha})$-space. In the latter case we need a $\bigO(\alpha2^{\alpha})$-time preprocessing phase in order to address the $2^{\alpha}$ possible registers. 

We are now ready to describe the three auxiliary algorithms used in the EPSM algorithm. The pseudocode of the three algorithms is shown in Fig. \ref{code:epsm}.

\subsection{\epsma: Searching for Very Short Patterns} \label{sec:new1}
The \epsma algorithm  is designed to be extremely fast in the case of very short patterns and 
although it could be adapted to work for longer patterns its performances degrades as the length of
the patterns increase. In practical cases the EPSM algorithm uses this procedure when $0<m< 4$.
The pseudocode of the algorithm is shown in Fig \ref{code:epsm}.

The preprocessing of the algorithm (lines 1-4) is computed on the prefix of the pattern of length $m' = \min\{m,\frac{\alpha}{2}\}$. If $m'=m$ the whole pattern is preprocessed and searched, otherwise the algorithm works as a filter, searching for all occurrences of the prefix with length $m'$ and, after an occurrence has been found, naively checking the whole occurrence of the pattern.

Specifically the preprocessing  phase consists in constructing an array $B$ of $m'$ different strings of length $\alpha$. Each string of the array exactly fits in a word of $w$ bits. The $i$-th string in the array $B$ consists of $\alpha$ copies of the character $\p_i$. More formally the string $B[i]$, for $0\leq i <m'$, is defined as $B[i]=(\p_i)^{\alpha}$.
For instance, if $\p=ab$ is a pattern of length $m=2$, $\gamma=8$ and $w=128$, then $B$ consists of two strings  of length $\alpha=16$, defined as 	$B[0] = a^{16}$  and $B[1] = b^{16}$.
The preprocessing phase of the algorithm requires $\bigO(\min\{m,\frac{\alpha}{2}\}\alpha)$-time and $\bigO(\min\{m,\frac{\alpha}{2}\})$-space.

The searching phase of the algorithm (lines 5-14) processes the text $\t$ in chunks of $\alpha$ characters. Let $N=\frac{n}{\alpha}-1$ and let $\tb=\tb_0\tb_1\ldots \tb_{N}$ be the string $\t$ represented in chunks of characters.
Each block of the text, $T_i$, is compared with the strings in the array $B$ using the instruction \textsf{wscmp}. 

Let $s_j=b_0b_1\ldots b_{\alpha-1}$ be the $\alpha$-bit register returned by  the instruction \textsf{wscmp}$(T_i, B[j])$, for $0\leq j<m'$.  It can be easily proved that $b_k=1$ if and only if the $k$-th character of the block $T_i$ is equal to $\p_j$, i.e. if and only if $T_i[k] = \p_j$ (remember that $B[j]=(\p_j)^{\alpha}$).
Finally let $r=r_0r_1\ldots r_{\alpha-1}$ be the $\alpha$-bit register defined as
$
	r = s_0\ \&\ (s_1 \ll 1)\ \&\ (s_2 \ll 2)\ \&\ \cdots\ \&\ (s_{m'-1} \ll (m'-1)).
$

It is easy to prove that $\p[0\pp m'-1]$ has an occurrence beginning at position $j$ of $T_i$ if and only if $r_j=1$. In fact $r_j=1$ only if $s_k[j+k]=1$, for $k=0\ldots m'-1$, which implies that $T_i[j+k] = \p_{k}$, for $k=0\ldots m'-1$.
Then, if $m=m'$ the algorithm reports the occurrences of the pattern at positions $i\alpha+\{r\}$, if any. Otherwise we know that occurrences of the prefix of the pattern with length $\alpha/2$ begin at positions $i\alpha+\{r\}$. Thus the algorithm checks the occurrences beginning at those positions. 

If we maintain, for each value $r$, with $0\leq r <2^{\alpha}$, a list of the values in the set $\{r\}$, the naive check of the occurrences can be done in $\bigO(|\{r\}|m)$-time.  When $m=m'$ the occurrences can be reported in $\bigO(|\{r\}|)$-time.
Finally, observe that the $m'-1$ possible occurrences crossing the blocks $T_i$ and $T_{i+1}$ are naively checked by the algorithm (lines 13-14).

The overall time complexity of the \epsma algorithm is $\bigO(nm)$, because in the worst case a naive check is required
for each position of the text. However, when $m\leq \frac{\alpha}{2}$ the \epsma algorithm achieves a $\bigO(n+occ)$
time complexity, where $occ$ is the number of occurrences of $\p$ in $\t$.



\subsection{\epsmb: Searching for Short Patterns} \label{sec:new2}
The \epsmb searches for the whole pattern when its length is less or equal to $\alpha/2$ and works as a
filter algorithm for longer patterns. However, it is based on a more efficient filtering technique and turns out to be
faster in the second case. 
In practical cases the EPSM algorithm uses this procedure when $m\geq 4$.
The pseudocode of the \epsmb algorithm is shown in Fig. \ref{code:epsm} (in the middle). Notice that no preprocessing phase is needed by the algorithm.

Let $m'$ be the minimum between $\alpha/2$ and $m$. Moreover  let $\p'$ be the prefix of $p$ of length $m'$.
The searching phase of the algorithm (lines 3-14) processes the text $\t$ in chunks of $\alpha$ characters. Let $N=\frac{n}{\alpha}-1$ and let $\tb=\tb_0\tb_1\ldots \tb_{N}$ be the string $\t$ represented in chunks of characters.
Each block of the text, $T_i$, is searched one by one for occurrences of the string $\p'$ using the instruction \textsf{wsmatch}. 
 
\begin{figure}[!t]
\begin{center}
\begin{scriptsize}
\begin{tabular}{|l|}
\hline
~~~~~~~~~~~~~~~~
\begin{tabular}{rl}
&\\
\multicolumn{2}{l}{~\textsc{\epsma($\p, m, \t, n$)}}\\
~\textsf{1.} & \textsf{$m' \leftarrow \min\{m,\alpha/2\}$}\\ 
~\textsf{2.} & \textsf{for $i \leftarrow 0$ to $(m'-1)$ do}\\ 
~\textsf{3.} & \qquad \textsf{for $j \leftarrow 0$  to $\alpha-1$ do}~\\ 
~\textsf{4.} & \qquad \qquad \textsf{$B_i[j] \leftarrow \p[i]$}\\ 
~\textsf{5.} & \textsf{for $i\leftarrow 0$ to $(n/\alpha)-1$ do}~\\ 
~\textsf{6.} & \qquad \textsf{$r \leftarrow 1^{\alpha}$}\\
~\textsf{7.} & \qquad \textsf{for $j \leftarrow 0$  to $m'-1$ do}\\
~\textsf{8.} & \qquad \qquad \textsf{$s_j \leftarrow$ wscmp$(T_i, B_j)$}\\
~\textsf{9.} & \qquad \qquad \textsf{$r \leftarrow r\ \&\ (s_j \ll j)$}\\
~\textsf{10.} & \qquad  \textsf{if $m=m'$}\\
~\textsf{11.} & \qquad \textsf{then report occurrences at $i\alpha+\{r\}$}~~\\
~\textsf{12.} & \qquad \textsf{else check positions $i\alpha+\{r\}$}\\
~\textsf{13.} & \qquad \textsf{for $j \leftarrow 0$  to $m-2$ do}\\
~\textsf{14.} & \qquad \qquad \textsf{check position  $(i+1)\alpha-j$}~\\
&\\
\multicolumn{2}{l}{~\textsc{\epsmb($\p, m, \t, n$)}}\\
~\textsf{1.} & \textsf{$m' \leftarrow \min\{m,\alpha/2\}$}\\ 
~\textsf{2.} & \textsf{$\p' \leftarrow \p[0 \pp m'-1]$}\\ 
~\textsf{3.} & \textsf{for $i\leftarrow 0$ to $(n/\alpha)-1$ do}~\\ 
~\textsf{4.} & \qquad \textsf{$r \leftarrow$ wsmatch$(\tb_i, \p')$}\\
~\textsf{5.} & \qquad \textsf{if $r \neq 0^{\alpha}$ then}\\
~\textsf{6.} & \qquad  \qquad \textsf{if $m=m'$}\\
~\textsf{7.} & \qquad \qquad \textsf{then report occurrences at $i\alpha+\{r\}$}~\\
~\textsf{8.} & \qquad \qquad \textsf{else check positions $i\alpha+\{r\}$}~\\
~\textsf{9.} & \qquad \textsf{$S \leftarrow$ wsblend$(\tb_i, \tb_{i+1})$}\\
~\textsf{10.} & \qquad \textsf{$r \leftarrow$ wsmatch$(S, \p')$}\\
~\textsf{11.} & \qquad \textsf{if $r \neq 0^{\alpha}$ then}\\
~\textsf{12.} & \qquad  \qquad \textsf{if $m=m'$}\\
~\textsf{13.} & \qquad \qquad \textsf{then report occurrences at $i\alpha+\frac{\alpha}{2}+\{r\}$}~~\\
~\textsf{14.} & \qquad \qquad \textsf{else check  positions $i\alpha+\frac{\alpha}{2}+\{r\}$}~\\
&\\
\multicolumn{2}{l}{~\textsc{\epsmc($\p, m, \t, n$)}}\\
~\textsf{1.} & \textsf{$mask \leftarrow 0^{\alpha-k}1^{k}$}\\
~\textsf{2.} & \textsf{for $i \leftarrow 1$ to $m-\alpha$ do }~\\
~\textsf{3.} & \qquad \textsf{$v \leftarrow \textsf{wscrc}(p[i..i+\alpha-1])$}\\
~\textsf{4.} & \qquad \textsf{$v \leftarrow v\ \&\ mask$}\\
~\textsf{5.} & \qquad \textsf{$L[v] \leftarrow L[v] \cup \{i\}$}\\
~\textsf{6.} & \textsf{$sh \leftarrow (\lfloor m/\alpha \rfloor -1)\cdot \alpha$}~\\
~\textsf{7.} & \textsf{for $i\leftarrow 0$ to $(n/\alpha)-1$ do}~~\\
~\textsf{8.} & \qquad \textsf{$v \leftarrow \textsf{wscrc}(\tb_i)$}~~\\
~\textsf{9.} & \qquad \textsf{$v \leftarrow v\ \&\ mask$}~~\\
~\textsf{10.} & \qquad \textsf{for all  $j \in L[v]$ do}\\
~\textsf{11.} & \qquad \qquad \textsf{if $0\leq i-j < n-m$ }\\
~\textsf{12.} & \qquad \qquad \textsf{then check position $i-j$}~~\\
~\textsf{13.} & \qquad \textsf{$i \leftarrow i + sh$}\\
&\\
&\\
\end{tabular}~~~~~~~~~~~~~~~~\\
\hline
\end{tabular}
\end{scriptsize}
\caption{The \epsma (on the top), the \epsmb (in the middle) and the \epsmc (on the bottom) auxiliary algorithms.}
\label{code:epsm}
\end{center}
\end{figure}

Specifically, let $r = r_0r_1\ldots r_{\alpha-1}$ be the $\alpha$-bit register returned by  the instruction \textsf{wsmatch}$(\tb_i, \p')$, for $0\leq j<m'$.  We have that $r_j = 1$ if and only if an occurrence of $\p'$ begins at positions $j$ of the block $\tb_i$, for $0\leq j < \alpha/2$. 
Then, if $m'=m$ (and hence $\p=\p'$) the algorithm simply returns positions $i\alpha+j$, such that $r_j=1$. Otherwise, if $m'<m$, the algorithm naively checks for the whole occurrences of the pattern starting at positions  $i\alpha+j$, such that $r_j=1$.

Notice that generally packed string matching instructions allow to read only blocks $\tb_i$ of $\alpha$ characters (128 bits in the case of SSE instructions), where $\tb_i=\t[i\alpha..(i+1)\alpha-1]$. 
Occurrences of the pattern beginning in the second half of the block $\tb_i$ are checked separately.
In particular a new block, $S$, obtained by applying the instruction \textsf{wsblend}$(\tb_i, \tb_{i+1})$, is processed
in a similar way as block $\tb_i$. In this case we report all occurrences of the pattern beginning at positions 
$i\alpha+\alpha/2+j$, with $0\leq j< \alpha/2$. One may argue that why blending is used instead of simply shifting the
window. The reason is the SSE instructions used in this context require the operands to be 16-byte
aligned in memory, where the performance degrades significantly otherwise. Thus, blending is more advantageous.

The resulting algorithm has an $\bigO(nm)$ worst case time complexity and require $\bigO(1)$ additional space.
When $m\leq \alpha/2$ the algorithm reaches the optimal $\bigO(n/\alpha + occ)$ worst case time complexity.

\subsection{\epsmc: Searching for Medium Length Patterns} \label{sec:new3}
The \epsmc algorithm  is designed to be faster for medium length patterns. 
It is based on a simple filtering method and uses a hash function for computing fingerprint values on blocks of $\alpha$
characters.
The fingerprint values are computed by using a hash function $h:\Sigma^{\alpha} \rightarrow \{0,1,\ldots, 2^{k}-1\}$, for a constant parameter $k$ (in practice we chose $k=11$).

The function $h$ is computed in a very fast way by using the \textsf{wscrc} specialized instruction, and in particular
$
	h(a) = \textsf{wscrc}(a)\ \&\ 0^{\alpha-k}1^{k}
$,
for each $x \in \Sigma^{\alpha}$.

The pseudocode of the \epsmc algorithm is shown in Fig. \ref{code:epsm}.

During the preprocessing phase (lines 1-6) a fingerprint values of $k$ bits is computed for all substrings of the pattern of length $\alpha$.
Then a table $L$ of size $2^k$ is computed in order to store starting positions of all substrings of the pattern, indexed by their fingerprint values. 
In particular we have
$
	L[v] = \{ i \ |\ h(p[i..i+\alpha-1]) = v \}
$,
for all $0\leq v < 2^k$.

The preprocessing phase of the \epsmc algorithm takes $\bigO(m+2^k)$-time.

Let $N=\frac{n}{\alpha}-1$ and let $\tb=\tb_0\tb_1\ldots \tb_{N}$ be the string $\t$ represented in chunks of characters.
During the searching phase (lines 7-13) the \epsmc algorithm inspects the blocks of the text in steps of $(\lfloor m/\alpha \rfloor -1)\cdot \alpha$ positions.
For each inspected block $\tb_i$ the fingerprint value $h(\tb_i)$ is computed and all positions in the set $\{ i\alpha - j\ |\ j \in F[h(\tb_i)]\}$ are naively checked.
The \epsmc algorithm has a $\bigO(nm)$ worst case time complexity but turns out to be very effective in practical cases.


\section{Experimental Results} \label{sec:results}
In this section we present experimental results in order to compare the performances of our newly presented algorithms against the best solutions known in literature in the case of short patterns.
We consider all the fastest algorithms in the case of short patterns as listed in a recent experimental evaluation by Faro and Lecroq \cite{faro13,faro10}. 
In particular  we compared the following algorithms:
\begin{itemize}

\item the Hash algorithm using groups of $q$ characters \cite{lecroq07} (HASH$q$);
\item the Extended Backward Oracle Matching algorithm \cite{faro08,faro09} (EBOM);
\item the TVSBS algorithm \cite{TVSBS} (TVSBS);
\item the Shift-Or algorithm  \cite{baezayates92} (SO)
\item the Shift-Or algorithm with $q$-grams \cite{durian09} (UFNDM$q$);
\item the Fast-Average-Optimal-Shift-Or algorithm \cite{FG05} (FAOSO$q$);
\item the Backward DAWG Matching algorithm using $q$-grams \cite{durian09} (BNDM$q$);
\item the Simplified BNDM$q$ algorithm \cite{durian09} (SBNDM$q$);
\item the Forward BNDM$q$ algorithm \cite{faro08,faro09,tarhio11} (FBNDM$q$);
\item the Crochemore-Perrin algorithm using SSE instructions \cite{benkiki11} (SSECP);
\item the EPSM algorithm presented in this paper.
\end{itemize}

We remember that the EPSM algorithm consists of the \epsma algorithm, when $m<4$, of the \epsmb algorithm when $4\leq m\leq 16$, and of the \epsmc algorithm when $m > 16$..
 
In the case of algorithms making use of $q$ grams, the value of $q$  ranges in the set $\{2,4,6\}$.
All algorithms have been implemented in the C programming language and have been tested using the Smart tool \cite{faro11} for exact string matching. The experiments were executed locally 
on a machine running Ubuntu 11.10 (oneiric) with Intel i7-2600 processor with 16GB memory.
Algorithms have been compared in terms of running times, including any preprocessing time.
For the evaluation we used a genome sequence, a protein sequence and a natural language text (English language), all sequences of 4MB. The sequences are provided by the Smart research tool. 
For each input file, we have searched sets of $1000$ patterns of fixed length $m$ randomly extracted from the text, for $m$ ranging from $2$ to $32$ (short patterns). Then, the mean of the running times has been reported. 

Table 1, Table 2 and Table 3 show the experimental results obtained for a gnome sequence, a protein sequence and a natural language text, respectively.

In the case of algorithms using $q$-grams  we have reported only the best result obtained by its variants. The values of
$q$ which obtained the best running times are reported as apices.
Running times are expressed in hundredths of seconds, best results have been boldfaced and underlined, while the second best results have been boldfaced.

\newcommand{\best}[1]{\underline{\textbf{#1}}}
\newcommand{\second}[1]{\textbf{#1}}

\begin{table}[!t]
\begin{scriptsize}
\renewcommand{\arraystretch}{1.3}
\begin{tabular*}{\textwidth}{@{\extracolsep{\fill}}|l|llllllllll|}
\hline
$m$ & $2$ & $4$ & $6$ & $8$ & $12$ & $16$ & $20$ & $24$ & $28$ & $32$\\
\hline
\textsc{HASH$q$} & - & 14.7$^{(3)}$ & 11.5$^{(3)}$ & 10.7$^{(3)}$ & 8.78$^{(3)}$ & 7.45$^{(3)}$ & 6.70$^{(3)}$ & 6.15$^{(5)}$ & 5.75$^{(5)}$ & 5.47$^{(5)}$\\
\textsc{EBOM} & 11.73 & 10.60 & 10.61 & 10.57 & 10.83 & 9.79 & 8.94 & 8.33 & 7.88 & 7.50\\
\textsc{TVSBS} & 16.17 & 13.78 & 12.60 & 11.93 & 11.29 & 10.90 & 10.74 & 10.61 & 10.56 & 10.45\\[0.1cm]
\hline
\textsc{SO} & 10.76 & 10.99 & 10.62 & 10.93 & 10.86 & 10.67 & 10.89 & 10.83 & 10.77 & 10.73\\
\textsc{FAOSO$q$} & - & 13.0$^{(2)}$ & 10.7$^{(2)}$ & 8.69$^{(2)}$ & 7.83$^{(2)}$ & 6.56$^{(4)}$ & 5.90$^{(4)}$ & 5.76$^{(4)}$ & 5.66$^{(4)}$ & 5.58$^{(4)}$\\
\textsc{UFNDM$q$} & 12.0$^{(2)}$ & 9.53$^{(4)}$ & 7.84$^{(4)}$ & 6.94$^{(4)}$ & 5.97$^{(6)}$ & 5.39$^{(6)}$ & 5.03$^{(6)}$ & 4.81$^{(6)}$ & 4.61$^{(6)}$ & 4.61$^{(6)}$\\
\textsc{BNDM$q$} & 12.8$^{(2)}$  & 11.3$^{(2)}$  & 9.23$^{(4)}$ & 7.24$^{(4)}$ & \best{5.90}$^{(4)}$ & 5.36$^{(4)}$ & 5.09$^{(4)}$ & 4.78$^{(6)}$ & 4.61$^{(6)}$ & 4.46$^{(6)}$\\
\textsc{SBNDM$q$} & 12.7$^{(1)}$ & 11.2$^{(2)}$ & 9.62$^{(4)}$ & 7.55$^{(4)}$ & 6.12$^{(4)}$ & 5.54$^{(4)}$ & 5.15$^{(6)}$ & 4.83$^{(6)}$ & 4.62$^{(6)}$ & 4.50$^{(6)}$\\
\textsc{FBNDM$q$} & 16.8 $^{(1)}$& 10.9$^{(4)}$ & 8.86$^{(4)}$ & 7.24$^{(4)}$ & 6.03$^{(4)}$ & 5.48$^{(4)}$ & 5.17$^{(6)}$ & 4.87$^{(6)}$ & 4.66$^{(6)}$ & 4.57$^{(6)}$\\[0.1cm]
\hline
\textsc{SSECP} & 5.31 & 5.59 &  \best{5.98} & 6.50 &  9.32 & 9.03 & 8.73 & 8.53 & 8.45 & 8.37\\
\textsc{EPSM} & \best{4.45} & \best{4.86} &6.18 & \best{6.12} & 6.16 &  \best{4.69} &  \best{4.77} &  \best{4.31} &  \best{4.38} &  \best{4.22}\\[0.1cm]
\hline
\end{tabular*}\\[0.1cm]
\end{scriptsize}
\label{tab:genome}
\caption{Experimental results for searching 1000 patterns on a genome sequence.}
\end{table}

\begin{table}[!t]
\begin{scriptsize}
\renewcommand{\arraystretch}{1.3}
\begin{tabular*}{\textwidth}{@{\extracolsep{\fill}}|l|llllllllll|}
\hline
$m$ & $2$ & $4$ & $6$ & $8$ & $12$ & $16$ & $20$ & $24$ & $28$ & $32$\\
\hline
\textsc{HASH$q$} & - & 14.1$^{(3)}$ & 11.3$^{(3)}$ & 11.2$^{(3)}$ & 8.28$^{(3)}$ & 6.98$^{(3)}$ & 6.29$^{(3)}$ & 5.81$^{(3)}$ & 5.51$^{(3)}$ & 5.27$^{(3)}$\\
\textsc{EBOM} & 10.00 & 6.25 & 5.50 & 5.14 & 4.84 & 4.69 & 4.60 & 4.58 & 4.53 & 4.51\\
\textsc{TVSBS} & 11.71 & 10.52 & 10.45 & 9.20 & 7.68 & 6.83 & 6.29 & 5.93 & 5.66 & 5.30\\[0.1cm]
\hline
\textsc{SO} & 10.68 & 10.68 & 10.67 & 10.62 & 10.67 & 10.76 & 10.70 & 10.51 & 10.69 & 10.21\\
\textsc{FAOSO$q$} & - & 8.54$^{(2)}$ & 7.82$^{(2)}$ & 6.42$^{(4)}$ & 5.70$^{(4)}$ & 5.67$^{(4)}$ & 5.15$^{(6)}$ & 5.12$^{(6)}$ & 5.10$^{(6)}$ & 5.09$^{(6)}$\\
\textsc{UFNDM$q$} & 11.0$^{(2)}$ & 7.69$^{(2)}$ & 6.44$^{(2)}$ & 5.80$^{(2)}$ & 5.18$^{(2)}$ & 4.85$^{(2)}$ & 4.62$^{(4)}$ & 4.46$^{(4)}$ & 4.33$^{(4)}$ & 4.22$^{(4)}$\\
\textsc{BNDM$q$} & 10.6$^{(2)}$ & 7.16$^{(2)}$ & 5.95$^{(2)}$ & 5.42$^{(2)}$ & 4.93$^{(2)}$ & 4.68$^{(2)}$ & \best{4.45}$^{(4)}$ & 4.30$^{(4)}$ & \best{4.18}$^{(4)}$ & \best{4.12}$^{(4)}$\\
\textsc{SBNDM$q$} & 10.4$^{(2)}$ & 7.01$^{(2)}$ & 5.86$^{(2)}$ & 5.37$^{(2)}$ & 4.89$^{(2)}$ & \best{4.65}$^{(2)}$ &  4.48$^{(4)}$ & 4.33$^{(4)}$ & 4.21$^{(4)}$ & 4.12$^{(4)}$\\
\textsc{FBNDM$q$} & 10.5$^{(1)}$ & 8.64$^{(1)}$ & 6.85$^{(2)}$ & 6.37$^{(4)}$ & 5.21$^{(4)}$ & 4.76$^{(4)}$ & 4.50$^{(4)}$ & 4.34$^{(4)}$ & 4.23$^{(4)}$ & 4.19$^{(4)}$\\[0.1cm]
\hline
\textsc{SSECP} & 5.31 & 5.58 & 5.96 & 6.49 & 6.68 & 6.45 & 6.33 & 6.24 & 6.19 & 6.15\\
\textsc{EPSM} & \best{4.47} & \best{4.83} & \best{4.65} & \best{4.65} & \best{4.64} & \best{4.65} & 4.73 & \best{4.28} & 4.31 & 4.18\\[0.1cm]
\hline
\end{tabular*}\\[0.1cm]
\end{scriptsize}
\label{tab:protein}
\caption{Experimental results for searching 1000 patterns on a protein sequence.}
\end{table}

\begin{table}[!t]
\begin{scriptsize}
\renewcommand{\arraystretch}{1.3}
\begin{tabular*}{\textwidth}{@{\extracolsep{\fill}}|l|llllllllll|}
\hline
$m$ & $2$ & $4$ & $6$ & $8$ & $12$ & $16$ & $20$ & $24$ & $28$ & $32$\\
\hline
\textsc{HASH$q$} & - & 14.2$^{(3)}$ & 11.2$^{(3)}$ & 11.1$^{(3)}$ & 8.29$^{(3)}$ & 6.99$^{(3)}$ & 6.26$^{(3)}$ & 5.83$^{(3)}$ & 5.50$^{(3)}$ & 5.27$^{(3)}$\\
\textsc{EBOM} & 10.26 & 7.24 & 6.51 & 6.14 & 5.82 & 5.67 & 5.55 & 5.53 & 5.42 & 5.37\\
\textsc{TVSBS} & 12.02 & 10.74 & 10.25 & 9.58 & 8.14 & 7.24 & 6.67 & 6.34 & 6.01 & 5.76\\[0.1cm]
\hline
\textsc{SO} & 10.87 & 10.80 & 10.63 & 10.72 & 10.72 & 10.79 & 10.59 & 10.71 & 10.66 & 10.72\\
\textsc{FAOSO$q$} & - & 9.22$^{(2)}$ & 8.01$^{(2)}$ & 6.89$^{(4)}$ & 5.77$^{(4)}$ & 5.66$^{(4)}$ & 5.20$^{(6)}$ & 5.10$^{(6)}$ & 5.11$^{(6)}$ & 5.10$^{(6)}$\\
\textsc{UFNDM$q$} & 10.6$^{(2)}$ & 8.33$^{(2)}$ & 7.20$^{(2)}$ & 6.35$^{(4)}$ & 5.48$^{(4)}$ & 5.02$^{(4)}$ & 4.77$^{(4)}$ & 4.62$^{(4)}$ & 4.47$^{(4)}$ & 4.39$^{(4)}$\\
\textsc{BNDM$q$} & 10.6$^{(2)}$ & 8.19$^{(2)}$ & 7.09$^{(2)}$ & 6.49$^{(2)}$ & 5.46$^{(4)}$ & 4.96$^{(4)}$ & 4.69$^{(4)}$ & 4.55$^{(4)}$ & 4.41$^{(4)}$ & 4.33$^{(4)}$\\
\textsc{SBNDM$q$} &10.8$^{(2)}$ & 8.02$^{(2)}$ & 6.99$^{(2)}$ & 6.44$^{(2)}$ & 5.60$^{(4)}$ & 5.07$^{(4)}$ & 4.78$^{(4)}$ & 4.64$^{(4)}$ & 4.48$^{(4)}$ & 4.39$^{(4)}$\\
\textsc{FBNDM$q$} & 10.5$^{(1)}$ & 9.13$^{(1)}$ & 8.07$^{(4)}$ & 6.65$^{(4)}$ & 5.51$^{(4)}$ & 5.02$^{(4)}$ & \best{4.74}$^{(4)}$& 4.61$^{(4)}$ & 4.45$^{(4)}$ & 4.43$^{(4)}$\\[0.1cm]
\hline
\textsc{SSECP} & 5.33 & 5.60 & 5.98 & 6.51 & 7.37 & 7.03 & 6.83 & 6.69 & 6.62 & 6.61\\
\textsc{EPSM} & \best{4.48} & \best{4.85} & \best{5.15} & \best{5.13} & \best{5.04} & \best{4.68} & 4.75 & \best{4.31} & \best{4.35} & \best{4.21}\\[0.1cm]
\hline
\end{tabular*}\\[0.1cm]
\end{scriptsize}
\label{tab:natural}
\caption{Experimental results for searching 1000 patterns on a natural language text.}
\end{table}

From experimental results it turns out that the EPSM algorithm has mostly the best performances for short patterns. When
searching on a genome sequence it is second only to the BNDM$q$ algorithm for $12\leq m\leq 14$ and to the SSECP
algorithm when $m=6$. Observe however that the EPSM algorithm is (up to 2 times) faster than the SSECP algorithm in most
cases.

When searching on a natural language text the EPSM algorithm obtains in most cases the best results, and is second to BNDM based algorithms only for $20 \leq m \leq 22$.

For increasing lengths of the pattern the performances of the EPSM algorithm remain stable, underling a linear trend on
average. However, the performances of other algorithms based on  shift heuristics, slightly increases.
This is more evident when searching on a protein sequence, where the algorithms based on bit-parallelism and $q$ grams 
turn out to be the faster solutions for longer patterns. 
However, in this latter cases the EPSM algorithm is always very close the best solutions.

It is interesting to observe that the EPSM algorithm is faster than the SSECP algorithm in almost all cases, and the gap
is more evident  in the case of longer patterns. In fact, despite to its optimal worst case time complexity, the SSECP
algorithm shows an increasing trend on average, while the EPSM algorithm shows a linear behavior.

\section{Conclusions} \label{sec:conclusions}
We presented a new packed exact string matching algorithm based on the Intel streaming SIMD extensions technology. The
presented algorithm, named EPSM, is based on three auxiliary algorithms which are used when $0<m< 4$, $m\geq 4$, and
$m \geq 16$, respectively. Despite the $\bigO(nm)$-worst case time complexity the resulting algorithm turns out to be
very fast in the case of very short patterns. From our experimental results it turns out that the EPSM algorithm is
in general the best solutions when $m \leq 32$. 
It could be interesting to investigate
the possibility to improve the performances of packed string matching algorithms by introducing shift heuristics.


\end{document}